\documentclass[12pt]{article}

\usepackage{graphics}
\usepackage{epsfig}
\usepackage{amssymb,amsmath}
 \makeatletter
\def\alt{\mathrel{\mathpalette\gl@align<}}
\def\agt{\mathrel{\mathpalette\gl@align>}}
\def\gl@align#1#2{\lower.6ex\vbox{\baselineskip\z@skip\lineskip\z@
\ialign{$\m@th#1\hfil##\hfil$\crcr#2\crcr\sim\crcr}}} \makeatother

\begin{document}

%

%
\vspace*{1.0cm}
\begin{center}
\baselineskip 20pt

{\Large\bf 
%
Direct and Indirect Detection and LHC Signals of Bino-Higgsino Dark Matter
}

\vspace{1cm}

{\large
Ilia Gogoladze$^a$\footnote{
Email: ilia@bartol.udel.edu. On leave of absence from:
Andronikashvili Institute of Physics, GAS, Tbilisi, Georgia.},
Rizwan Khalid$^a$\footnote{
Email: rizwan.hep@gmail.com. On study leave from:
Centre for Advanced Mathematics \& Physics of the National
University of Sciences \& Technology, H-12, Islamabad, Pakistan. },
Yukihiro Mimura$^{a,\,b}$,
%
and
Qaisar Shafi$^a$}

\vspace{1.0cm}

{
\it
$^a$Bartol Research Institute, Department of Physics and Astronomy, \\
University of Delaware, Newark, DE 19716, USA 

$^b$Department of Physics, National Taiwan University, Taipei,
Taiwan 10617, R.O.C.
}

\vspace{.5cm}

\vspace{1.5cm} {\bf Abstract}
\end{center}

\vspace{.5cm}


If the lightest dark matter neutralino has a sufficiently large Higgsino 
component, its spin-independent and spin-dependent cross sections on nucleons 
can be sizable enough to be detected soon in direct and indirect surveys.
We outline in this paper some characteristic features expected of mixed 
bino-Higgsino dark matter. 
If the observed relic density is saturated by the bino-Higgsino dark matter,
it fixes the amount of allowable bino-Higgsino mixing
and provides predictions for other observables which can be tested at the
Large Hadron Collider (LHC).
We study the correlation between the cross sections 
and the branching ratio of $B_s\to \mu^+\mu^-$. 
For a mixed bino-Higgsino dark matter, 
the mass differences of the neutralinos can be less than $M_Z$.
This will cause an excess of lepton pairs, above the Standard Model 
predictions, from the decays of the two heavier neutralinos. 
We discuss implications of 
the dilepton invariant mass distribution,
and outline a way to 
extract the neutralino parameters 
for testing gaugino mass unification
and deducing the relic density from an
interplay of astrophysical detection and LHC measurements.


\noindent

\thispagestyle{empty}
\newpage


\baselineskip 18pt


\section{Introduction}

There exists overwhelming evidence, most recently from 
the Wilkinson Microwave Anisotropy Probe (WMAP) \cite{Komatsu:2008hk}, 
that non-baryonic cold dark matter comprises 
around 23 percent of the Universe's 
energy density.
Identifying this dark matter, presumably an elementary particle,
is one of the foremost contemporary challenges in particle 
physics and cosmology. 
The goals for successful identification of dark matter are:
(1) Detection of the relic dark matter particle,
and measurement of its mass and distribution directly.
(2) Production of the dark matter particle at the LHC and future linear colliders,
and measurement of its properties.
(3) Testing the consistency between these measurements, namely in 
astrophysics and particle physics,
and reproduction of the relic abundance of the particle from the measured properties
in order to confirm that the dark matter 
particle (possibly more than one species of particles) 
really makes up 23 percent of the Universe's energy density. 

One of the most compelling features of low scale supersymmetry (SUSY), 
supplemented with $R$-parity conservation, is that it can provide an
attractive cold dark matter candidate with the correct relic abundance,
provided the lightest neutralino $\tilde\chi_1$
is also the lightest SUSY particle (LSP) \cite{DMreview}. 
If the LSP neutralino is bino dominated 
(in an admixture of bino, wino, and Higgsinos),  
it often leads to an over-abundance of dark matter,
unless (co)annihilation processes reduce the relic density 
to levels compatible with WMAP.
Many solutions have been proposed to accomplish 
this \cite{DMreview,silkphysrep}.

One attractive
scenario for realizing the correct relic abundance is to consider an
appropriate bino-Higgsino mixture in the composition of the
LSP \cite{focus, hyper, funnel}. 
In this mixed bino-Higgsino LSP (called bino-Higgsino dark matter) scenario,
two neutralinos and one chargino have masses that are close to 
the LSP mass, such that (co)annihilation processes among them can 
reproduce the desired relic density.
%
The spin-independent (SI) cross section on nuclei in this scenario
is enhanced, which is an advantage from the point of view of direct detection
experiments searching for the LSP. 
Indeed, the recent 
candidate events reported by CDMSII \cite{Ahmed:2009zw} 
and EDELWEISS-II \cite{Edelweiss} would suggest that
the SI cross section is $O(10^{-8})\,{\rm pb} $. 
This is of the right order of magnitude 
for the bino-Higgsino dark matter scenario \cite{CDMSlightWIMP,CDMSotherSUSY}. 
As the bounds on the cross section get lowered by the ongoing and planned  
measurements by XENON100 \cite{Aprile:2009zzc}, SuperCDMS \cite{Schnee:2005pj}, 
and XMASS \cite{Abe:2009zz}, 
the WMAP compatible bino-Higgsino mixing solutions will be among 
the first ones to be tested.
%
%
Moreover, it is known that a significant Higgsino component 
in the LSP neutralino also gives
a large spin-dependent (SD) cross section,
which would make the indirect detection of this dark matter 
through self-annihilation into neutrinos and other particles 
more feasible.
Thus, the bino-Higgsino dark matter solution 
will be tested by the IceCube/Deep Core neutrino observatory~\cite{Hultqvist:2010xy}.
It is important to observe both SI and SD cross 
sections and to see their correlation \cite{Cohen:2010gj} 
in order to adequately test the
bino-Higgsino dark matter scenario.


The SI cross section is enhanced 
if the mass $m_A$ of the CP-odd Higgs  boson 
is small and $\tan\beta$ (ratio
of Higgs vacuum expectation values (VEVs)) is large.
With $m_A$ small, pair annihilation processes are enhanced,
and a reduced bino-Higgsino mixing can 
give rise to the desired WMAP relic density.
We refer to this case as bino-Higgsino-like dark matter,
if we need to distinguish among the WMAP solutions.
On the other hand, when we specify a mixed bino-Higgsino LSP solution
where (co)annihilation processes via scalars are negligible, 
we refer to it as well-tempered bino-Higgsino dark matter \cite{Masiero:2004ft}.
{}From the particle physics point of view,
the rare decay $B_s\to\mu^+\mu^-$ is one of the most 
interesting processes in the region of large
$\tan\beta$ and small $m_A$ \cite{Choudhury:1998ze}.
%
The Tevatron will provide a bound ($\sim 2\times10^{-8}$) 
on this branching ratio in run II \cite{CDF}, and LHCb, within a few years, will probe 
the standard model prediction $(3-4)\times 10^{-9}$. 
(The exclusion limit from 1 fb$^{-1}$ of data expected by the end of 2011
will be $\sim 6 \times 10^{-9}$)
\cite{Lenzi:2007nq}.
It is thus important to investigate the regions of parameter space 
that provide larger SI cross sections
(small $m_A$ and/or small Higgsino mass $\mu$),
and to explore their predictions.

At the LHC, the neutralino LSP is created 
from cascade decays of squarks and gluinos,
and manifests itself as missing energy.
As mentioned above, to identify dark matter, one major goal is to
reproduce the LSP relic density from the collider measurements \cite{Feng:2001ce}.
%
%
However, the inverse problem at the LHC \cite{ArkaniHamed:2005px}
is not so easy in general, 
since 
it is hard to measure the mass spectrum and the couplings directly.
Several techniques have been developed, 
and several reliable relic density simulations have been explored  
for various WMAP solutions 
\cite{Polesello:2004qy,Arnowitt:2006jq,Nath:2010zj,Dutta:2010uk}.
The assumptions of universality 
and/or unification of the SUSY breaking mass parameters
are crucial simplifications for the collider measurements of the relic density.
For the bino-Higgsino dark matter,
universality of the sfermion masses is less important
since, by definition, the coannihilation processes via sfermions are negligible
as far as the relic density is concerned.
The gaugino mass spectrum 
(which should also be addressed at the LHC \cite{Altunkaynak:2009tg})
will be more important in restricting the relic density with LHC measurements.

%

%

%

%

In this paper we investigate bino-Higgsino dark matter and its
implications for direct and indirect detection,
and for the LHC measurements.
%
%
We will study both the well-tempered bino-Higgsino dark matter
and 
bino-Higgsino-like dark matter with smaller $m_A$.
In the study of the well-tempered mixing solution, 
it is assumed that the sfermions are
sufficiently heavy, without specifying the SUSY breaking scenario or
any underlying theory.
This is done in order to make  
the bino-Higgsino dark matter relic density,
and the SI and SD cross sections insensitive to these masses.
The bino-Higgsino mixing needed to satisfy the desired WMAP relic density
depends on wino and bino mass ratio,
and thus the cross sections implicitly depend on this ratio.
For simplicity, we assume gaugino unification at 
the grand unification scale, $M_{\rm GUT}$, 
in order to investigate the cross sections.
We also study the possibility of non-universal
gauginos, which also can be tested experimentally within our
framework. 
Our results should be applicable to any model where
the well-tempered bino-Higgsino dark matter solution can be realized. 
On the other hand, in order to exhibit 
our study of the correlation of cross sections
and Br($B_s \to \mu^+\mu^-$) for bino-Higgsino-like dark matter,
we employ non-universal Higgs mass boundary conditions,
where $m_A$ and $\mu$ are free low energy parameters.
%
In this case, for a given LSP mass and $m_A$, 
the proper WMAP relic density constrains the Higgsino mass $\mu$.
As a result, the chargino contribution to Br($B_s \to \mu^+\mu^-$)
is predictable for a given stop mass, if
gaugino mass unification is assumed.


In our presentation we first study the constraints and implications
from SI and SD cross sections
for the bino-Higgsino(-like) dark matter solution.
If the SI cross section is large ($\sigma_{\rm SI} \gtrsim 10^{-8}$ pb), 
the bino-Higgsino mixing is large and/or $m_A$ is small. 
%
%
The SD cross section is restricted, for given $m_A$
if the bino-Higgsino mixing is determined
by the WMAP observation.
%
%
If the CP-odd Higgs mass $m_A$ is small, the
amount of bino-Higgsino mixing required to satisfy
the WMAP relic density is not very large. 
The SD cross section, accordingly, is then also not very large. 
Hence it is worth making
clear the conditions under which
we can observe the SD cross section
by indirect detection,
as well as the corresponding prediction for 
Br$(B_s\to\mu^+\mu^-)$, while
satisfying the other experimental constraints.
When $m_A$ is large, the bino-Higgsino mixing needs to be well-tempered
and the SD cross section must be large.
We also investigate the bound on SD cross section for smaller
neutralino masses, $\lesssim 100$ GeV, since it is already bounded by the
recent CDMSII / XENON100 data.

We then proceed to study the implication from LHC measurements.
If the bino-Higgsino mixing is well-tempered,
three of the mass eigenvalues of the neutralino mass matrix
can be degenerate to within $O(M_Z)$,
depending on the neutralino mass parameters.
In such a case, the dilepton invariant mass distribution
from the heavier neutralino decay with missing energy
will give us important information on the neutralino mass parameters.
Due to the large SI cross section, the mass of the bino-Higgsino 
dark matter particle is expected to be measured
from the distribution of the recoil energy
of the heavy nuclei in the direct detection experiments, and
there arises a possibility to extract the parameters
for reproducing the LSP relic density and the gaugino mass spectrum.


This article is organized as follows. 
In Section 2, 
the SI and SD cross sections of
neutralino-nucleon interactions are briefly studied.
In Section 3, the correlation between
the bino-Higgsino dark matter solution and 
SI cross section is described. 
Within the bino-Higgsino dark matter scenario, the interplay
between the SD cross section and 
Br$(B_s\rightarrow\mu^+\mu^-)$ is presented in
Section 4. 
We discuss in Section 5 several possible signatures of
this scenario at the LHC,
and in Section 6 we summarize our results.

\section{Spin-Independent and Spin-Dependent Cross Sections}

The Higgs exchange diagrams dominate the SI cross
section of the lightest neutralino on nucleon \cite{Jungman:1995df} as long as
squarks are sufficiently heavy.
Also, for $m_H \lesssim m_h \sqrt{\tan\beta}$, 
the contribution from the heavier Higgs ($H$) exchange is
dominant over the lighter Higgs ($h$).
The SI cross section in this case  can be written as
\begin{equation}
\sigma_{\rm SI} \simeq
\frac{m_N^4}{4\pi}\frac{g_2^4}{M_W^2} \frac{\cos^2\alpha}{\cos^2\beta}
\frac{F_H^2}{m_H^4}
 \left[(f_d + f_s + \frac{2}{27} f_G) + \frac{\tan\alpha}{\tan\beta}
 (f_u + \frac{4}{27} f_G)\right]^2,
\label{eq-SI}
\end{equation}
where $\alpha$ is the Higgs mixing angle, $\tan\beta\equiv
\langle H_2 \rangle / \langle H_1\rangle$ is the ratio of up- and down-type Higgs 
VEVs, $M_W$ is the $W$ boson mass, $g_2$ is the $SU(2)$ gauge
coupling, $m_N$ is the nucleon mass,  
$f_q = m_q \langle N | \bar q q | N \rangle/ m_N$ for nucleon $N$,
$f_G = 1- f_u - f_d - f_s$, and
$F_H =(N_{12}-N_{11}\tan\theta_W) (N_{14} \sin\alpha - N_{13}\cos\alpha)$. 
$N_{1i}$ are the elements of the diagonalizing matrix of the neutralino mass matrix
such that the lightest neutralino can be written as a linear combination
of gauginos and Higgsinos:
\begin{equation}
\tilde\chi_{1}= N_{11} \tilde B + N_{12} \tilde W + N_{13} \tilde H_1
+ N_{14} \tilde H_2.
\end{equation}
%
%
The lighter Higgs exchange also contributes to the
SI cross section which, therefore, does not vanish ($\sim 10^{-8}$ pb)
even if $m_A \gtrsim 1$ TeV
in the case of bino-Higgsino dark matter ($N_{11} N_{13} \sim 0.1$).


The strange sea quark content of the
nucleon is very important from the point of view of computing the SI cross section.
Recent lattice collaborations report small values of $f_s$
\cite{Ohki:2008ff,Young:2009zb,Toussaint:2009pz}.
The smallest value is reported by the JLQCD collaboration in 2009 as $f_s = 0.02$, 
with $f_s < 0.08$ to within $1\sigma$.
For $f_s = 0.118$ (which is the default value of the numerical package 
ISAJET \cite{Baer:1999sp} that we use), 
the cross section is roughly a factor 2 larger when compared to the case
of small $f_s$. 
%
If we use a larger value of $f_s$ ($\sim 0.2-0.4$),
the bino-Higgsino dark matter is on the edge of the
current bound set by CDMSII and XENON100,
and is even excluded in particular for small $m_H$.
We will use the value $f_s = 0.03$ in this paper.
For $f_{u,d}$, the default values of ISAJET are used
($f_u= 0.023, f_d = 0.034$ for protons).

The SD cross section, $\sigma_{\rm SD}$,
is dominated by the $Z$ boson exchange diagram \cite{DMreview,Jungman:1995df}
provided the squarks are sufficiently heavy,
\begin{equation}
\sigma_{\rm SD} \propto \frac{1}{M_Z^4} (N_{13}^2- N_{14}^2)^2.
\end{equation}
In this case, the SD cross section depends only on
neutralino mixing (disregarding hadronic uncertainty),
and it thus provides a good probe of the gaugino and Higgsino parameters.
%
Large SD cross sections clearly prefer a large Higgsino component for the
LSP.
If the LSP mass is less than $M_Z$, the SD cross section is
maximized while satisfying the WMAP data. However, if we assume
gaugino mass unification, the bino-Higgsino solution with a large SD
cross section is excluded by the chargino mass bound
$m_{\tilde\chi_1^+} \geq 103$ GeV \cite{PDG}.

%

As the dark matter gets scattered by nucleons in the sun it loses
kinetic energy, and eventually will not be able to escape from the sun's gravity.
As a result,
a larger SD cross section gives rise to a 
larger population of neutralinos 
in and around the sun \cite{DMreview}.
This population of neutralinos will manifest itself 
through self-annihilation into 
high energy neutrinos (or muons) emerging from the sun.
%
%
%
High energy neutrinos can be observed
if the SD cross section is sufficiently large \cite{Ellis:2009ka}.
The bino-Higgsino mixing solution was explored by
AMANDA \cite{Ackermann:2005fr},
and it will also be tested by IceCube/Deep Core \cite{Hultqvist:2010xy}.

\section{Bino-Higgsino Dark Matter and SI Cross Section}

In the bino-Higgsino dark matter scenario, 
the SI cross section is almost determined by the parameters $m_A$ and $\mu$ 
for a given LSP mass and $\tan\beta$. For a more general description of
this scenario, we consider non-universal Higgs mass boundary
conditions so as to keep $\mu$ and $m_A$ free. 
We assume universal trilinear couplings, $A_0$, 
as well as universal soft gaugino and sfermion masses $m_{1/2}$ 
and $m_0$ respectively.
For the purposes of this analysis, the universality of
sfermion masses is not crucial,
especially for the case of well-tempered bino-Higgsino dark matter.

\begin{figure}[t]
\center
\includegraphics[width=8cm]{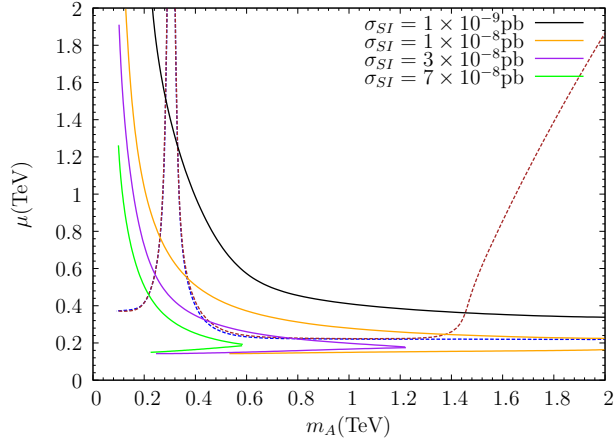}
\caption{
The contours for SI cross section (solid lines)
and $\Omega h^2 = 0.11$ (dotted lines) for a
fixed lightest neutralino mass $m_{\tilde\chi_1} = 150$ GeV.
The $\Omega h^2 \sim 0.11$ contour is shown for
$m_0=500 \,{\rm GeV}$
(dotted red line) and $m_0=2$ TeV (dotted blue line).
The SI cross section contours correspond to $1\times 10^{-9} {\rm pb}$ (black line),
$1\times 10^{-8} {\rm pb}$ (orange line), $3\times 10^{-8} {\rm pb}$ (purple line) and
$7\times 10^{-8} {\rm pb}$ (green line).
}
\label{Fig1}
\end{figure}

In Fig.~\ref{Fig1} we show contours for $\Omega h^2$ and $\sigma_{\rm SI}$
in the $m_A$-$\mu$ plane for fixed lightest neutralino mass $m_{\tilde\chi_1} = 150$ GeV,
$\tan\beta=40$ and $A_0=0$. The $\Omega h^2 \sim 0.11$ contour is shown for 
$m_0=500 \,{\rm GeV}$
(dotted red line) and $m_0=2$ TeV (dotted blue line).
The SI cross section contours correspond to $1\times 10^{-9} {\rm pb}$ (black line),
$1\times 10^{-8} {\rm pb}$ (orange line), $3\times 10^{-8} {\rm pb}$ (purple line) and
$7\times 10^{-8} {\rm pb}$ (green line). For $m_A = 2 m_{\tilde\chi_1}$,
$\Omega h^2$ is tiny ($\ll 0.1$) due to the $s$-channel resonance in the
pair annihilation processes,
and the so called $A$-funnel WMAP solutions lie near this line. For smaller $\mu$,
the second lightest neutralino gets closer in mass to $m_{\tilde\chi_1}$
and bino-Higgsino mixing becomes larger.
Therefore, the WMAP contour shifts away from the $A$-resonance solution.
For $m_A \lesssim 2 m_{\tilde \chi_1} - m_h$,
the pair annihilation channels
$\tilde \chi_1 \tilde \chi_1 \to H h, Z A$
open up. As a result, the bino-Higgsino mixing should be smaller
($\mu$ should be larger) for the left branch of the $\Omega h^2$ solutions,
as shown in the figure.

Next let us describe the behavior of sfermion masses in the $m_A$-$\mu$ plane
for fixed $m_0$, $\tan\beta$ and $A_0$.
With smaller $\mu$, the SUSY breaking up-type Higgs mass squared 
is larger (but not its absolute value since $m_{H_u}^2 \simeq M_Z^2/2 - \mu^2$).
Therefore, the RGE evolution of sfermion masses, which couple to up-type Higgs,
drives the masses to smaller values.
An exception occurs if Higgsino dominates the lightest neutralino,
$\mu \sim m_{\tilde\chi_1}$.
In this region, $M_1$ is larger (for fixed $m_{\tilde\chi_1}$),
and the sfermion masses are also larger
due to the wino and gluino loops since gaugino unification is assumed.
For larger $m_A$ and/or smaller $\mu$,
the SUSY breaking down-type Higgs squared mass
is larger
because $m_A^2 = 2\mu^2 +m_{H_u}^2 + m_{H_d}^2$ and, thus,
$m_{H_d}^2 \simeq m_A^2 -\mu^2 - M_Z^2/2$.
The RGE evolution for sfermions masses, which couple to the down-type Higgs, therefore,
drives them to smaller values, especially if $\tan\beta$ is large.
As a result, 
for small $m_0$,
the stau-coannihilation region (and also the stau LSP region)
appears in the bottom-right corner of Fig.~\ref{Fig1},
and the $\Omega h^2$ contour is thus lifted up.
Furthermore, in the region of small $\mu$,
$m_{H_u}^2 > m_{H_d}^2$ is needed at the unification scale.
This then forces the sfermions with positive hypercharges to become lighter.
In the region of large $\mu$, on the other hand,
the sfermions with negative hypercharges become lighter,
and thus a sneutrino can be NLSP for large $\mu$ and small $m_0$, 
such that sneutrino-coannihilation can be realized.
Since we are interested in smaller $\mu$ values to obtain
large cross sections, we do not discuss the
coannihilation solutions.

In order to facilitate direct detection and determination of the
dark matter mass,
we are interested in large SI cross sections
$\gtrsim 10^{-8}$ pb.
One can see from Fig.~\ref{Fig1} that
the points that satisfy both the WMAP data and $\sigma_{\rm SI} > 10^{-8}$ pb
are separated into two regions because of the $A$-resonance.
One region corresponds to $m_A < 2 m_{\tilde\chi_1}$ (left branch), 
while the other to $m_A > 2 m_{\tilde\chi_1}$ (right branch).
The SD cross section is larger for the region $m_A > 2 m_{\tilde\chi_1}$
because of the larger bino-Higgsino mixing needed to satisfy WMAP
data, as previously described.
Such large SD cross sections are definitely testable at
IceCube/Deep Core \cite{Hultqvist:2010xy}.
For the $m_A < 2 m_{\tilde\chi_1}$ region,
the SD cross section is correspondingly smaller
to satisfy WMAP data.
However, in this region, the branching ratio of the decay 
$B_s\to \mu^+\mu^-$ can be large (for $\tan\beta \gtrsim 30$)
as the heavier Higgs can be light
and the Higgs penguin contribution is enhanced.
The SUSY enhancement of $B_s\to\mu^+\mu^-$ can be tested at the LHC,
and it is important to investigate the
possible values of the branching ratio.
Also, in this region,
the SD cross section may be on the verge of being
tested by IceCube/Deep Core.
%

\section{Correlation between Cross Sections and Br$(B_s \to \mu^+\mu^-)$}

Since the SI cross section and Br$(B_s\to\mu^+\mu^-)$ are sensitive to
the heavier Higgs mass,
a numerical analysis to see the correlation between the two is
interesting \cite{Ellis:2006jy}.
If there exists non-minimal flavor violation,
the CP violating phase in $B_s$-$\bar B_s$ mixing
can also be important \cite{Dutta:2009iy}.
In this section, we will
explore the prospects of neutralino detection via
the SD cross section inferred from observation of neutrino flux
from the sun by IceCube/Deep core observations. We will also discuss
the correlation between the SD cross section and the SUSY contribution 
to Br$(B_s\to\mu^+\mu^-)$ in the case of minimal flavor violation.


As mentioned in the previous section,
we employ non-universal Higgs boundary conditions
with universal sfermion masses and gaugino mass unification
in order to exhibit our results.
Since universality of sfermion masses is not crucial
to describe the bino-Higgsino mixing solutions, 
we do not impose constraints from the slepton mass spectrum
(arising from muon $g-2$ for instance).
Instead, 
we employ the
constraints from $b\to s\gamma$ 
to describe the correlation between SI and SD cross sections
and Br$(B_s\to\mu^+\mu^-)$.
%
%


If no FCNC source is introduced in the SUSY breaking mass parameters,
the important contribution to Br$(b\to s\gamma)$
comes from the chargino and charged Higgs loops.
The chargino contribution to the amplitude for $b\to s\gamma$
transition 
is (naively) proportional to $\tan\beta$,
while the charged Higgs contribution does not
depend very much on $\tan\beta$.
The latter contribution 
has a positive sign for the amplitude, 
while the chargino contribution gives a negative sign for the amplitude
when $\mu>0$.
These two contributions
can therefore be canceled through
an appropriate choice of parameters.  

For $m_A > 400-500$ GeV (with a slight dependence on $\tan\beta$),
the branching fraction Br($b\to s\gamma$) is smaller than $\sim 4.2\times 10^{-4}$
in the SUSY particle decoupling limit.
Therefore, only lower bounds on the SUSY particle masses are obtained
in this case.
If the Higgsino mass is fixed to obtain the proper relic density
for a given LSP mass, the
stop mass is bounded from below.
%

For $m_A < 400-500$ GeV, however,
the chargino contribution is needed to satisfy the experimental constraint on
Br$(b\to s\gamma)$,
and the stop mass is also bounded from above.
%
%
For small $\tan\beta$, in particular, 
the LEPII bound $m_h > 114.4$ GeV
can be more important than Br$(b\to s\gamma)$ for the lower bound
on the stop mass. 
As a result, small $m_A$ values can be
excluded by a combination of Br$(b\to s\gamma)$ and
$m_h$ bounds.

\begin{figure}[t]
\center
\includegraphics[width=0.45 \columnwidth=0.2]{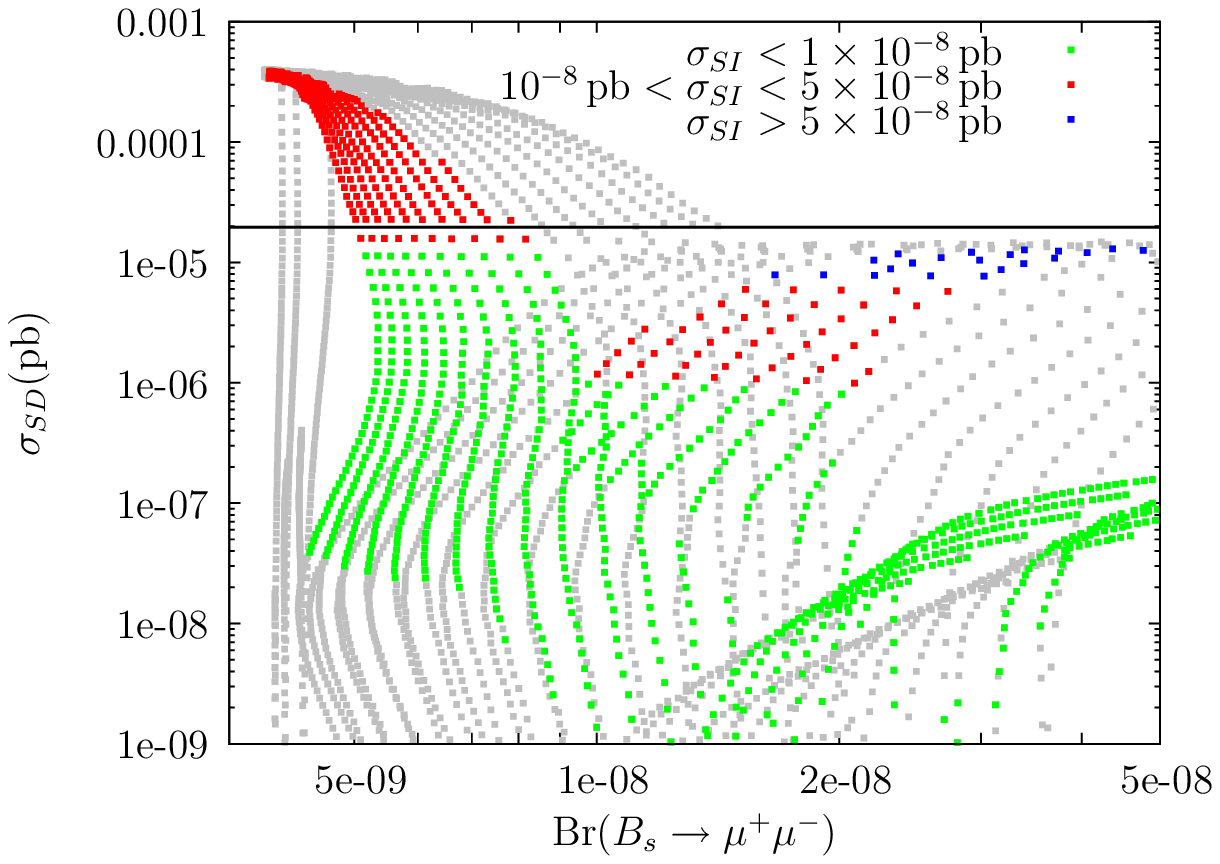}
\includegraphics[width=0.45 \columnwidth=0.2]{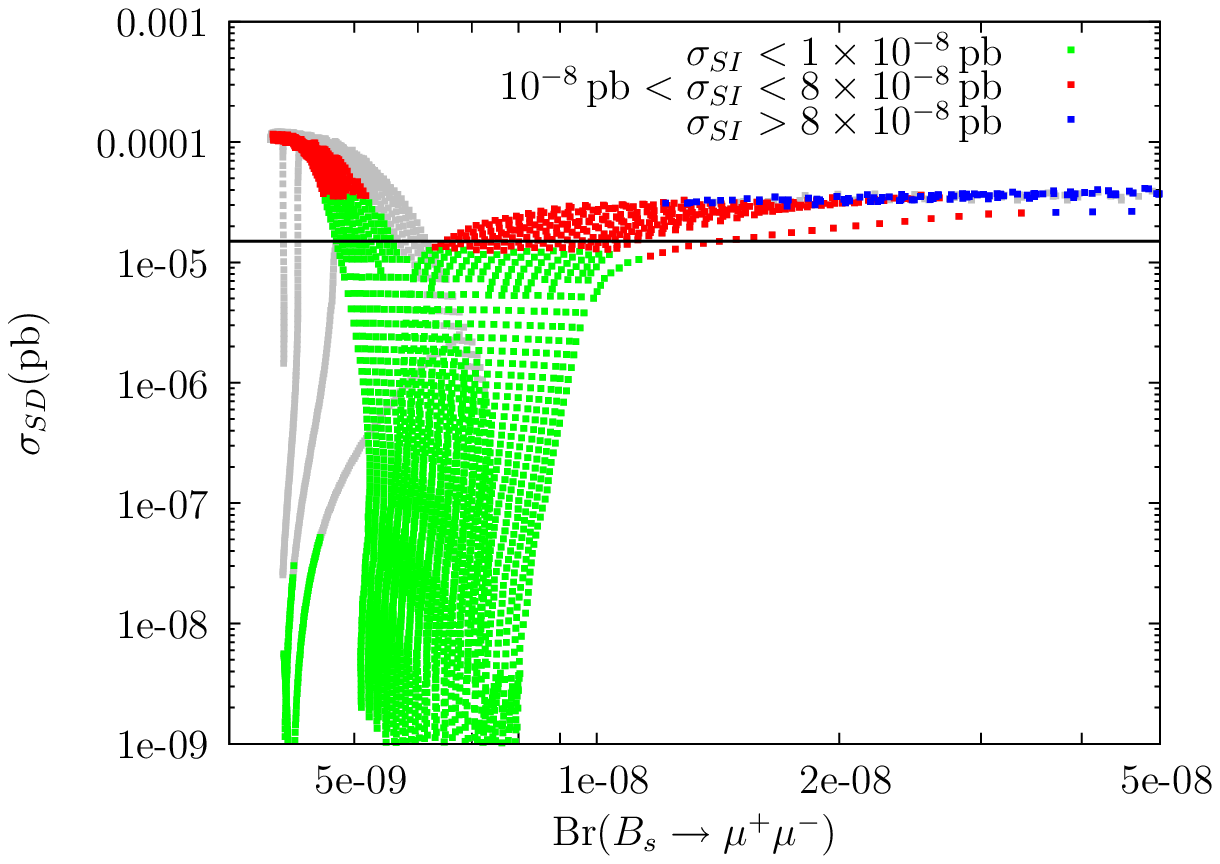}
\caption{
Br($B_s\to\mu\mu$) vs $\sigma_{\rm SD}$ plots
for $\Omega h^2= 0.11$,
$\tan\beta=40$, $A_0 = 0$, $m_{\tilde\chi_1} = 150$ GeV (left),
$m_{\tilde\chi_1} = 300$ GeV (right).
Green points satisfy Br($b\to s\gamma$) and $m_h$ bounds.
Red points satisfy the CDMSII bound and $\sigma_{\rm SI}> 10^{-8}$ pb.
Blue points are excluded by the CDMSII bound.
The current bound on Br$(B_s\to\mu^+\mu^-)$ is $4.3 \times 10^{-8}$
at the 95\% confidence level.
The horizontal lines indicate the expected sensitivity
of IceCube/Deep Core.
}
\label{Fig2}
\end{figure}

Fig. 2 shows a plot in Br$(B_s \to \mu^+\mu^-)$ - $\sigma_{\rm SD}$
plane
for $m_{\tilde\chi_1} = 150$ GeV (left panel) and 300 GeV (right panel).
The points shown satisfy the WMAP $2\sigma$ bounds on $\Omega h^2$ 
and are generated using
$m_0 < 2$ TeV, $0< \mu < 2$ TeV, $m_A < 2$ TeV,
$A_0 =0$ and $\tan\beta = 40$.
In the right branch ($m_A > 2 m_{\chi_1}$) of the WMAP solution,
$m_A$ is large and thus Br$(B_s \to \mu^+\mu^-)$ is 
comparable to the SM prediction.
One finds that the SD cross section is large in this branch.
%
The points for the left branch ($m_A < 2 m_{\chi_1}$) have 
Br$(B_s\to\mu^+\mu^-)$ that is bounded from both 
above and below.
The maximal values of Br$(B_s\to\mu^+\mu^-)$
and the SD cross section in the left branch have
already been excluded by the CDMSII bound.

The SD cross section for the left branch
is below the sensitivity of IceCube/Deep core
for $m_{\tilde\chi_1}=150$ GeV,
but lies on the boundary for 
$m_{\tilde\chi_1}= 300\, {\rm GeV}$.
Indeed, to observe both a large Br$(B_s \to\mu^+\mu^-)$, 
enhanced by sufficiently small $m_A$, and a large SD cross section
in the left branch solution,
the neutralino should be heavier than about 300 GeV
in order to satisfy the WMAP data.

We note that Br$(B_s\to\mu^+\mu^-)$
can be large and comparable to the current bound (even if $\tan\beta=30$)
when $\mu$ is large, of order $\sim 2$ TeV ($A$-funnel solution).
This is because of the finite correction from gluino loop contribution,
which generates a $b$-$s$ flavor changing Higgs coupling.
Although the gluino FCNC is suppressed, the contribution can be
large, for large $\mu$, due to the large left-right sbottom mixing. 
The cross sections are certainly small for the large $\mu$ solution.
The corresponding points can be seen in the case of $m_{\chi_1} = 150$ GeV
at the bottom-right side in Fig. 2 (left panel).

We should mention that
the left branch solution is disfavored
especially if $m_{\tilde \chi_1}$ is small and $\tan\beta$ is large,
since the charged Higgs mass (more precisely, $m_{H^+}^2/\tan\beta$)
is bounded
by the $B \to \tau\bar\nu$ and $D \to \tau\bar\nu$
constraints \cite{Hou:1992sy,Akeroyd:2009tn}.
However, the bounds are sensitive to the quark mixing parameters,
and so we keep the left branch solution for the WMAP relic density.
The SD cross section is not sensitive to $\tan\beta$
(if $\tan\beta \gtrsim 20$),
and thus there is a possibility to observe a large Br$(B_s \to\mu^+\mu^-)$
at LHCb
and a large SD cross section at the IceCube/Deep Core,
if $\tan\beta \sim 30$
and $m_{\tilde \chi_1} \gtrsim 300$ GeV.

\begin{figure}[t]
\center
\includegraphics[width=10cm]{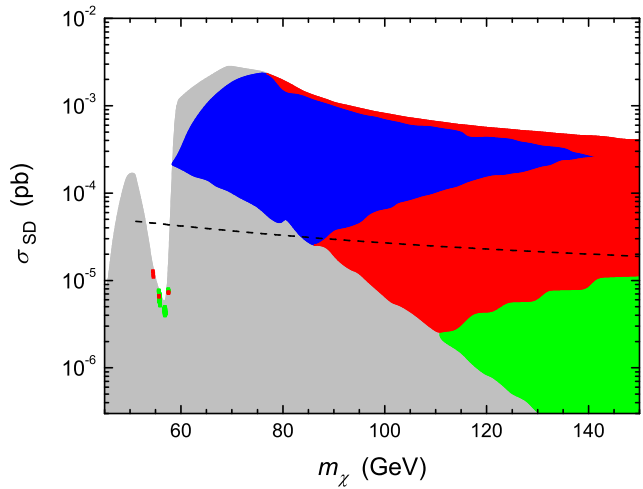}
\caption{
SD cross section vs neutralino mass
for $\Omega h^2= 0.11$,
$\tan\beta=40$, $m_0 = 2$ TeV, $A_0=0$.
We choose $f_s = 0.03$.
The red region is allowed by CDMSII and XENON100 and $\sigma_{\rm SI} > 10^{-8}$ pb.
The blue region is excluded by the CDMSII and XENON100 bounds.
For 80 GeV $\lesssim m_{\tilde\chi_1} \lesssim  130$ GeV,
the region just below the maximal SD cross section is excluded
(depending on $f_s$).
The dotted line indicates the
expected sensitivity of IceCube-80/Deep Core (1800d) \cite{Hultqvist:2010xy}.
%
%
%
}
\label{Fig3}
\end{figure}

Since $m_A$ can be small in the left branch,
the bound from direct detection
can exclude the blue points in Fig. 2.
If $m_{\tilde\chi_1}$ is small ($\lesssim 100$ GeV),
a region from the right branch of the WMAP solution can even 
be excluded.

In Fig. 3,
we plot the correlation between the SD cross section and the 
neutralino mass in the right branch solution for $\Omega h^2 =0.11$.
We find that a region
just below the maximal values of the possible SD cross section
is excluded by the CDMSII experiment, for 80 GeV $\lesssim m_{\tilde\chi_1}
\lesssim 130$ GeV.
Along the WMAP solution for the right branch (shown in Fig. 1),
the SI cross section increases for smaller $\mu$ values.
However, since $m_A$ is larger in this direction,
the SI cross section becomes maximal
and then decreases after that, while $\mu$ asymptotes to a minimum.
As a result, the region
just below the maximal SD cross section is excluded
if the maximal value of the SI cross section
is larger than the CDMSII bound.

{}From Fig. 3 one can see that $m_{\tilde \chi_1} \lesssim 80$ GeV
is excluded.
The exact numerical value depends on the chosen $f_s$ value.
For example,
if $f_s = 0.118$ (ISAJET default) is used,
the maximal SD cross section is excluded for $m_{\tilde\chi_1} \lesssim 120$ GeV.

\section{LHC phenomenology}

As previously mentioned, the
bino-Higgsino mixing needs to be well-tempered, 
especially
if $m_A \gg 2 m_{\tilde \chi_1}$
and sfermions are heavy.
In this case,
since the bino and Higgsino masses $M_1$ and $\mu$
need to be close together, the
three eigenvalues of the neutralino mass matrix
are approximately degenerate:
\begin{equation}
m_{\tilde\chi_2} - m_{\tilde\chi_1},\
m_{\tilde\chi_3} - m_{\tilde\chi_1} < M_Z.
\end{equation}
The second and third lightest neutralinos ($\tilde\chi_2$, $\tilde\chi_3$),
produced from the decays of squarks and/or gluino,
themselves decay into $\tilde\chi_1 \ell^+ \ell^-$.
The end points of the dilepton invariant mass, $M_{\ell\ell}$,
gives the mass differences of the neutralinos \cite{Baer:1986vf}.
If the mass differences are less than about 80 GeV,
we can measure two end points of the $M_{\ell\ell}$ distribution,
and this can yield important information
about the neutralino mass parameters.
In order to measure the mass differences,
the end points should be a little less than the $Z$ boson mass 
in order to avoid a $Z$-pole of the distribution.

The neutralino mass matrix is commonly written as
\begin{equation}
M_\chi =
\left(
 \begin{array}{cccc}
   M_1 & 0 & -M_Z \cos\beta \sin\theta_W & M_Z \sin\beta \sin\theta_W \\
   0   & M_2 & M_Z \cos\beta \cos\theta_W & -M_Z \sin\beta \cos\theta_W \\
   -M_Z \cos\beta \sin\theta_W & M_Z \cos\beta \cos\theta_W& 0 & -\mu \\
   M_Z \sin\beta \sin\theta_W& -M_Z \sin\beta \cos\theta_W & -\mu & 0
 \end{array}
\right).
\end{equation}
Because there are only off-diagonal entries in the Higgsino block,
the second and third mass eigenvalues are of
opposite signature
in the case of bino-Higgsino dark matter.
The relative sign of the neutralino mass is physical
for the $M_{\ell\ell}$ distribution.
In the limit where $m_{\tilde\chi_i} \ll m_{\tilde \ell}$,
the differential decay width of $\tilde\chi_i \to \tilde\chi_1 \ell\ell$
is \cite{Nojiri:1999ki,Kitano:2006gv,De Sanctis:2007td}
\begin{eqnarray}
\frac{d\Gamma}{dM_{\ell\ell}}
\propto
\frac{M_{\ell\ell}}{(M_{\ell\ell}^2-M_Z^2)^2}
\sqrt{\left( (m_{\tilde\chi_i} - m_{\tilde\chi_1})^2 - M_{\ell\ell}^2\right)
\left((m_{\tilde\chi_i} + m_{\tilde\chi_1})^2 - M_{\ell\ell}^2 \right)} \\
\times
\left((\eta_i m_{\tilde\chi_i} - m_{\tilde\chi_1})^2 + 2 M_{\ell\ell}^2 \right)
\left((\eta_i m_{\tilde\chi_i} + m_{\tilde\chi_1})^2 -M_{\ell\ell}^2 \right).
\nonumber
\end{eqnarray}
%
As a convention, 
all $m_{\tilde\chi_i}$ take positive values,
and the eigenstate $\tilde \chi_1$ is assigned a positive mass eigenvalue.
If the remaining eigenstates
$\tilde\chi_i$, $i=2,3,4$, have a positive (negative) eigenvalue,
we define the corresponding $\eta_i = 1$ $(-1)$.
%
%

It is easy to see that in the limit where
$m_{\tilde\chi_i}- m_{\tilde\chi_1} \ll M_Z$,
the distribution is almost symmetric for $\eta_i = -1$, 
and the peak of the distribution is at half of the end point
$M_{\ell\ell}^{\rm end} = m_{\tilde\chi_i} - m_{\tilde\chi_1}$.
Due to the factor
${M_{\ell\ell}}/{(M_{\ell\ell}^2-M_Z^2)^2}$,
the distribution near the end point is enhanced
when the mass difference is close to $M_Z$.
For $\eta_i = 1$,
the peak shifts towards the end point
even if the mass difference is not close to $M_Z$.
Therefore, by
observing the shape of the $M_{\ell\ell}$ distribution,
one can distinguish between $\eta_i$ positive or negative.

We denote the eigenstate where $\eta_i = 1$ $(-1)$
as $\tilde\chi_p$ ($\tilde\chi_m$).
Then, by definition,
$\tilde\chi_2 = \tilde \chi_p$
and $\tilde\chi_3 = \tilde \chi_m$
if $m_{\tilde\chi_p} < m_{\tilde \chi_m}$, for example.

We have the following four equations among the MSSM parameters
and the eigenvalues:
\begin{eqnarray}
&&m_{\tilde\chi_1} + m_{\tilde\chi_p} - m_{\tilde\chi_m} + m_{\tilde\chi_4} =M_1+M_2, 
\label{eq1}
\\
&&m_{\tilde\chi_1}^2 + m_{\tilde\chi_p}^2 + m_{\tilde\chi_m}^2 + m_{\tilde\chi_4}^2
   =M_1^2+M_2^2+2(\mu^2+M_Z^2), 
\label{eq2}
\\
&&m_{\tilde\chi_1}^3 + m_{\tilde\chi_p}^3 - m_{\tilde\chi_m}^3 + m_{\tilde\chi_4}^3
   =M_1^3+M_2^3+ 3 (M_1 \sin^2\theta_W+M_2 \cos^2\theta_W + \mu \sin2\beta)M_Z^2, 
\label{eq3}
\\
&&-m_{\tilde\chi_1} m_{\tilde\chi_p} m_{\tilde\chi_m} m_{\tilde\chi_4}
   =-\mu^2 M_1 M_2 + \mu M_Z^2 (M_1 \cos^2\theta_W + M_2 \sin^2\theta_W)\sin2\beta.
\label{eq4}
\end{eqnarray}
Suppose that the mass differences, $D_p \equiv m_{\tilde \chi_p} - m_{\tilde \chi_1}$
and $D_m \equiv m_{\tilde \chi_m} - m_{\tilde \chi_1}$,
are accurately measured at the LHC (which can be done to an accuracy of
$\pm 1$ GeV \cite{Tovey}),
and there remain six unknown parameters:
$m_{\tilde \chi_1}$, $m_{\tilde\chi_4}$, $M_1$, $M_2$, $\mu$ and $\tan\beta$.
If gaugino unification is assumed,
the ratio $M_2/M_1$ is almost fixed at low energy
in the neutralino mass matrix. We can then
solve the equation as a function of $\tan\beta$.

\begin{figure}[t]
\center
\includegraphics[width=8cm]{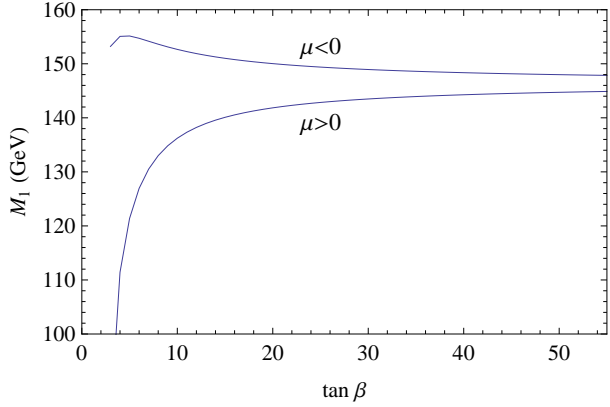}
\caption{
Plot of $M_1$ vs. $\tan\beta$
for $D_p = m_{\tilde\chi_p}-m_{\tilde\chi_1} = 60$ GeV
and $D_m = m_{\tilde\chi_m}-m_{\tilde\chi_1} = 80$ GeV.
Gaugino mass unification is assumed, $M_2/M_1 \simeq 2$.
}
\label{Fig4}
\end{figure}

Assuming that $D_p$ and $D_m$ are measured to be $60$ GeV and 
$80$ GeV respectively at the LHC,
we plot the solution for $M_1$ as a function of $\tan\beta$
in Fig. 4.
The solution is less sensitive
to $\tan\beta$ if $\tan\beta \gtrsim 20$,
which is reasonable from the
form of the neutralino mass matrix
and the fact that when $M_1 \sim \mu$, 
$M_Z^2 \sin2\beta$ can be neglected in the equations
for large $\tan\beta$. 
To solve the equations, we assume that the $\mu$ parameter is real.

If squarks are much heavier than the gluino (as in the focus
point/hyperbolic branch solution in the constrained MSSM (CMSSM) \cite{Chan:1997bi}),
the gluino mass can be measured (to within an accuracy of 10\%)
\cite{Baer:2007ya}. We can then determine whether or not we have gaugino mass 
unification for $M_1$ and $M_3$. In the example of Fig. 4, if
$M_3$ is measured to be about 900 GeV, we may conclude that 
there exists nice unification of gaugino masses ($M_1$ and $M_3$) for
$\tan\beta \gtrsim 20$. If $M_3$ is less than about 900 GeV, 
unification is still possible for $\tan\beta \lesssim 10$.

A model-independent measurement of $\tan\beta$ is important
to conclude whether we have gaugino mass unification
from the measurement of neutralino mass differences.
It is hard to determine $\tan\beta$ at the LHC
model-independently,
but it can be measured by a future $e^+ e^-$ linear collider \cite{Berggren:1999sr}.

As previously mentioned, 
for large $\tan\beta$ 
the solution for $M_1$ (as shown in Fig. 4) 
becomes less sensitive to $\tan\beta$. 
In this case, $m_{\chi_1}$ can also be restricted
from the measurement of the dilepton invariant mass distribution. 
The SI cross section is large, $\sim 10^{-8}$ pb, for the 
well-tempered bino-Higgsino LSP, and
thus it can be expected that $m_{\tilde\chi_1}$ is measured by
the distribution of recoil energy in direct detection
experiments. 
If the LSP mass is accurately measured, we can determine
whether $\tan\beta$ is small ($\lesssim 10$) or not,
by comparing the restriction from the measurements of neutralino mass differences. 
To do this, however, we need to assume gaugino mass unification for
$M_1$ and $M_2$.

\begin{figure}[t]
\center
\includegraphics[width=8cm]{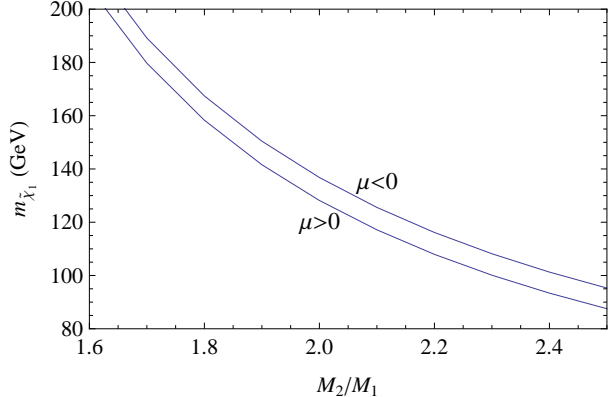}
\caption{
The lightest neutralino mass $m_{\tilde\chi_1}$ is plotted
as a function of the gaugino mass ratio $M_2/M_1$
for $\tan\beta =30$, $D_p = 60$, and $D_m = 80$ GeV.
}
\label{Fig5}
\end{figure}

In Fig. 5,
we plot the lightest neutralino mass $m_{\tilde\chi_1}$
as a function
of the ratio $M_2/M_1$, with
$\tan\beta=30$,
$D_p = 60$ GeV and $D_m = 80$ GeV.
If it turns out that $m_{\tilde\chi_1}$
is larger than about 140 GeV
from the direct detection experiments,
the ratio $M_2/M_1$ needs to be smaller than 2
($M_2/M_1 \simeq 2$ is the expectation from gaugino mass unification).
In this case, bino-wino-Higgsino tri-mixing may be realized.
If $m_{\tilde\chi_1}$ is less than 140 GeV,
it is possible that
either $M_2/M_1$ is larger than 2
or gaugino unification is realized for $\tan\beta \lesssim 10$.
Once one knows that $\tan\beta$ is large ($\gtrsim 20$) from 
other experiments, for instance,
one can conclude that $M_2/M_1$ is larger than 2.
%
Therefore, an independent measurement of $\tan\beta$
will be important in order to test gaugino unification
from the $D_p$ and $D_m$ measurements.

\begin{figure}[t]
\center
\includegraphics[width=8cm]{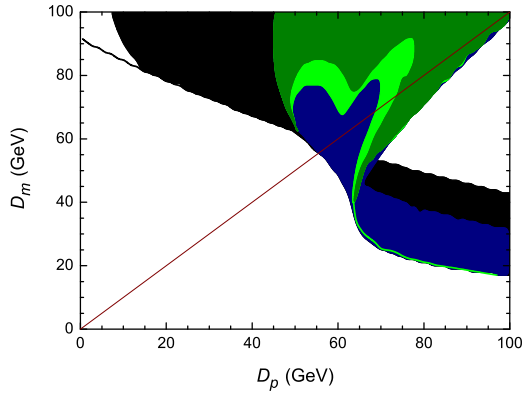}
\caption{
Map of the solution for given mass differences, $D_p$, $D_m$
in the case where gaugino mass is unified, $M_2/M_1 \simeq 2$.
The black colored region is excluded by the chargino mass bound.
The other color codes correspond to varying
relic density $\Omega h^2$, as explained in the text.
%
}
\label{Fig6}
\end{figure}

In Fig. 6, we plot the solutions of Eqs.(\ref{eq1})-(\ref{eq4})
for $D_p, D_m < 100$ GeV, 
assuming $\tan\beta = 30$, $\mu> 0$
and gaugino mass unification, $M_2/M_1 \simeq 2$.
In the black region,
there are solutions for given $D_p, D_m$,
but the chargino mass bound, $m_{\tilde\chi_1^+}> 103$ GeV,
 is not satisfied.
The colored regions blue, light and dark green,
satisfy the chargino mass bound.

The $M_{\ell\ell}$ distribution, of course, is independent of whether
the WMAP relic density is provided for or not. We present different
color codings for varying WMAP relic density in the case
where sfermion and heavier Higgs masses are 2 TeV to make the
bino-Higgsino mixing well-tempered. We show $\Omega h^2 < 0.085$
in blue,  $0.085 < \Omega h^2 < 0.13$ in light green, and 
$\Omega h^2 > 0.13$ in dark green color.
%
The focus point/hyperbolic branch solutions in the CMSSM
should lie in the light green region.
However, if $\tan\beta \sim 50$, the heavier Higgs can be light 
which enhances the neutralino annihilation cross section. 
As a result, the focus point/hyperbolic branch of the CMSSM will 
penetrate into the dark green region of Fig. 6. 
It is not necessary to have the unification condition
for SUSY breaking scalar masses
since the neutralino mass differences are independent of this assumption.
If coannihilation with a sfermion is present,
the relic density is reduced and the dark green region can satisfy
the WMAP data.

The reason that the shape of WMAP solutions resembles a heart is
as follows: For the left ventricle, the lightest
neutralino mass is less than about 170 GeV. 
(Therefore, the SD cross section is larger for the left side, $D_p \simeq 45-60$ GeV). 
For the right ventricle, $\tilde \chi_1$ is heavier than 170 GeV, and so the
neutralinos can pair-annihilate to a top pair, and thus
smaller bino-Higgsino mixing is needed, and $D_m$ becomes larger. 
%
In bino-Higgsino dark matter, a larger Higgsino 
component is required for a heavier neutralino 
in the absence of coannihilations with scalar particles. 
So the heavier the neutralino, 
the closer $\mu$ needs to be to $M_1$. 
Therefore, $D_m$ and $D_p$ decrease for larger $m_{\chi_1}$ $(\gtrsim 200$ GeV),
which forms the right ventricle.
For $m_{\tilde \chi_1} \gtrsim 300$ GeV, the ordering of mass eigenvalues
of $\tilde\chi_p$ and $\tilde\chi_m$ is flipped since $\mu$ is
closer to $M_1$, 
namely $D_m < D_p$ for $m_{\tilde \chi_1} \gtrsim 300$ GeV. 
The light green line for $D_m < 40$ GeV corresponds to a
Higgsino-like LSP ($\mu < M_1$), with $m_{\tilde \chi_1} \sim 1$
TeV.

Except for $m_{\tilde\chi_1} \sim 200$ GeV,
the SI cross section can be $10^{-8}$ pb
even if $m_A = 2$ TeV.
In the case of $m_{\tilde\chi_1} \sim 200$ GeV,
$D_m$ is close to $M_Z$ to satisfy the neutralino relic density.
It may then be difficult to measure it due to the $Z$-pole
of the $M_{\ell\ell}$ distribution.
For $\tan\beta \lesssim 10$, $D_m$ may be larger
and closer to $M_Z$ for the WMAP solution,
even for $m_{\tilde \chi_1}\sim 100$ GeV.
The SD and SI cross sections can be large even in this case. 
If the cross sections are experimentally observed, the LSP is
bino-Higgsino dark matter and one of the mass differences can
restrict the parameter space.

It is interesting that $D_p < D_m$ is satisfied for 
not too heavy LSP ($m_{\tilde\chi_1}\lesssim 300$ GeV), while for a relatively
heavy LSP, the opposite holds, $D_m < D_p$. 
This just follows from the neutralino mass matrix  
when $M_1$ and $M_2$ have the same sign for the well-tempered
bino-Higgsino dark matter.
%
%
If $M_1$ and $M_2$ have opposite signs, 
on the other hand,
$D_m < D_p$  
as long as the chargino mass bound is satisfied and $D_p < 100$ GeV. 
In Fig. 7, we plot the solutions for $M_2=-2M_1$ and $\tan\beta=30$.

\begin{figure}[t]
\center
\includegraphics[width=8cm]{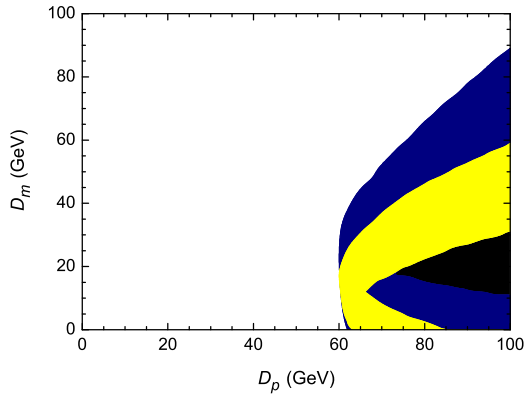}
\caption{
Map of the solutions for given mass differences
in the case of $M_2/M_1 = -2$.
Black region is excluded by the chargino mass bound.
Yellow region satisfies $\sigma_{\rm SD} > 10^{-4}$ pb.
It is important that there is no point for $D_m > D_p$,
which means that the eigenvalues of
$\tilde\chi_1$ and $\tilde\chi_2$ have opposite signs.
}
\label{Fig7}
\end{figure}

%

Because the $M_{\ell\ell}$ distribution looks very different,
it is a powerful tool to observe the relative signatures of 
$M_1$ and $M_2$
if the lightest neutralino is not heavy,  $m_{\tilde \chi_1} \lesssim 300$ GeV.
If $M_1 M_2 >0$ for the relatively light LSP,
$D_p < D_m$ is satisfied.
The distribution near the end point
is then enhanced,
and it has a sharp edge near $M_{\ell\ell} \sim D_p$.
If the edge is not enhanced and it turns out that the LSP is light,
it follows that $M_1 M_2 <0$ and
gaugino mass unification is not realized.

We note that the edge can be enhanced even if $D_m < D_p$
is satisfied for $M_1 M_2 < 0$
due to the factor
$M_{\ell\ell}/(M_{\ell\ell}^2-M_Z^2)^2$ in the distribution function.
This can happen when $D_m$ is close to $M_Z$, which is not the case for 
the light neutralino. 
For example, in Fig. 7
we show the region where the SD cross section is more than
$10^{-4}$ pb.
The region corresponds to a light neutralino,
and $D_p$ is less than 50 GeV.
In this case, the edge of the distribution cannot be enhanced.

Finally, it is also possible for 
$\tilde\chi_2$ and $\tilde \chi_3$ to have mass eigenvalues
with the same signature if $|M_2/M_1| \sim 1$.

\section{Summary}

We have investigated the direct and indirect detection of the
bino-Higgsino dark matter scenario, and explored 
its implications for the LHC.
%
%
We first presented the prediction of SD cross section and 
the branching ratio of $B_s \to \mu^+\mu^-$, assuming that
the WMAP relic density constraint is satisfied. 
Because the
relic density restricts the bino-Higgsino mixing, the SD cross
section is predicted for given $m_A$. 
For the WMAP compatible solution, we have
two regions for bino-Higgsino  dark matter : (1) $m_A < 2 m_{\tilde
\chi_1}$, (2) $m_A > 2 m_{\tilde \chi_1}$. 
Since pair annihilation
channels can open up in region (1), the bino-Higgsino mixing here should be
smaller than in region (2). 
In region
(1), the SD cross section is therefore smaller, and the neutralino should be
sufficiently heavy for the SD cross section to be observed. 
We find
that the SD cross section can be observed indirectly by the neutrino
flux from the sun if $m_{\tilde \chi_1} \gtrsim 300$ GeV. 
In region~(1), the branching ratio for $B_s\to\mu^+\mu^-$ can be enhanced, and 
Br$(b\to s\gamma)$ constraint gives a lower bound in flavor universal
models which can be tested at LHCb. 
The CDMSII bound can exclude
a large branching ratio of $B_s\to\mu^+\mu^-$.
In region~(2), on the other hand, 
the bino-Higgsino mixing is well-tempered and the
SD cross section is large enough to be observed. For a neutralino
mass less than about 100 GeV, the CDMSII bound constrains the SD
cross section even in region~(2). 
A SD cross section just below
the maximal value, for given neutralino mass, is excluded by the
CDMSII experiment. 
This exclusion depends on the strange
sea-quark content $f_s$ in the nucleon (multiplied by the strange mass).
If the maximal SD cross section for $m_{\tilde\chi_1} \simeq
80-100$ GeV is observed, a smaller value ($f_s\sim0.03$) 
consistent with the recent results from lattice calculations 
will be preferred. 

We next studied the LHC phenomenology of bino-Higgsino dark matter.
Because the mass differences of the neutralinos in the case of
well-tempered bino-Higgsino dark matter are small, they can
be measured by the dilepton invariant mass distribution. 
From the
neutralino mass differences, we may be able to infer whether
gaugino masses are unified or not.
For this, it turns out that $\tan\beta$ is an important parameter.
If we find that $\tan\beta$ is large, say from an observation
such as Br$(B_s\to\mu^+\mu^-)$,
the gaugino mass ratio at the weak scale
can be obtained from the mass differences.
The shape of the dilepton invariant mass distribution
depends on the relative signatures of the neutralino mass eigenvalues. 
This distribution will be a powerful tool
in providing important information about neutralino masses
and the relative signatures of the gaugino masses.

If gaugino mass unification is assumed
and two of the mass differences of the neutralinos are measured,
$\tan\beta$ can be determined,
and the bino-Higgsino dark matter relic abundance 
is then reproduced.
The relic density thus deduced from  
collider measurements provides a strong hint for identifying the 
nature of dark matter if it coincides with the WMAP data. 
If the two do not coincide, 
we cannot decide whether
gaugino unification is not satisfied
or the neutralino LSP alone does not saturate the WMAP measured relic abundance.
A model-independent measurement of $\tan\beta$
provides a strong hint to solve this dilemma.
In general, it is hard to measure $\tan\beta$ model-independently
at the LHC, but it is possible at a future linear collider.
The polarization of $\tau$ lepton may give us a hint
of the size of $\tan\beta$
if sleptons are light enough
in the bino-Higgsino dark matter scenario \cite{Nojiri:1994it}.
We have in this paper assumed that all the sfermions are
heavy in order to make their mass parameters insensitive to our discussion,
but this assumption can be relaxed.
The large bino-Higgsino mixing can provide various features
for collider phenomenology, such as  
$\tau$ polarization,
if on-shell sleptons appear in the cascade decays.

From a theoretical point of view,
a confirmation of the bino-Higgsino dark matter
scenario can provide important impetus to investigations
of SUSY breaking.
A bino-Higgsino dark matter needs a relatively small
Higgsino mass $\mu$.
In fact, small $\mu$ is preferable
if it is a parameter independent of the 
SUSY breaking scale,
while $\mu$ can be large among the electroweak symmetry breaking vacua if
it depends on a single SUSY breaking scale parameter \cite{Giudice:2006sn}.
Therefore, testing the bino-Higgsino dark matter scenario can serve as an important
avenue for distinguishing among the various models of SUSY breaking.

%

In conclusion,
the lightest neutralino 
with an appropriate composition of bino and Higgsino components
is a compelling dark matter candidate.
This will soon be tested by
the ongoing and planned direct detection experiments,
and indirectly at the IceCube neutrino telescope through
pair annihilation.
A mixed bino-Higgsino dark matter particle can also 
lead to characteristic signals at the LHC as we have discussed.

\section*{Acknowledgments}

We thank B.~Dutta for useful discussions.
This work
is supported in part by the DOE Grant No. DE-FG02-91ER40626
(I.G., R.K., Y.M., and Q.S.) and GNSF Grant No. 07\_462\_4-270 (I.G.).
The work of Y.M. is supported by the Excellent Research Projects of
National Taiwan University under grant number NTU-98R0526.

\end{document}